\documentclass[a4paper, superscriptaddress, notitlepage, amsmath,amssymb, aps, twocolumn, prx, floatfix, 10pt]{revtex4-2}
\usepackage{graphicx}
\usepackage{bm,bbm}
\usepackage[urlcolor=blue]{hyperref}
\hypersetup{colorlinks=true,allcolors=blue}
\usepackage{amssymb} 

\newcommand{\E}[1]{\langle #1\rangle}

\newcommand{\rmd}{d}
\newcommand{\rme}{\mathrm{e}}
\newcommand{\rmi}{\mathrm{i}}
\newcommand{\tr}{\mathrm{tr}}
\newcommand{\f}[1]{\mathbf{#1}}
\newcommand{\x}{\f x}

\newcommand{\om}{\omega}
\newcommand{\abs}[1]{\left\lvert #1 \right\rvert}

\newcommand{\norm}[1]{\lvert\lvert #1 \rvert\rvert}

\newcommand{\bsig}{\boldsymbol{\sigma}}
\newcommand{\bSig}{\boldsymbol{\Sigma}}
\newcommand{\balpha}{\boldsymbol{\alpha}}
\newcommand{\bbeta}{\boldsymbol{\beta}}

\newcommand{\bgam}{\boldsymbol{\gamma}}

\newcommand{\ps}{p_{\rm s}}

\newcommand{\N}{\mathbb{N}}
\newcommand{\R}{\mathbb{R}}
\newcommand{\C}{\mathbb{C}}

\newcommand{\T}{{\tilde T}}

\usepackage{color}\usepackage{xcolor}
\colorlet{mylinkcolor}{blue!66!black!80}
\definecolor{grey}{rgb}{0.6,0.6,.6}
\definecolor{darkgrey}{rgb}{0.4,0.4,.4}
\definecolor{darkgreen}{rgb}{0,0.4,0}
\definecolor{lightgreen}{rgb}{0,0.7,0}
\definecolor{darkred}{rgb}{0.5,0,0}

\newcommand{\blue}[1]{{\color{black}#1}}

\begin{document}
\title{Asymmetric Thermal Relaxation in
  Driven Systems: Rotations go Opposite Ways}
\author{Cai Dieball}
\affiliation{Mathematical bioPhysics group, Max Planck Institute for Multidisciplinary Sciences, G\"{o}ttingen 37077, Germany}

\author{Gerrit Wellecke}
\affiliation{Mathematical bioPhysics group, Max Planck Institute for Multidisciplinary Sciences, G\"{o}ttingen 37077, Germany}
\affiliation{Present address: Theory of Biological Fluids, Max Planck Institute for Dynamics and Self-Organization, G\"{o}ttingen 37077, Germany}

\author{Alja\v{z} Godec}
\email{agodec@mpinat.mpg.de}
\affiliation{Mathematical bioPhysics group, Max Planck Institute for Multidisciplinary Sciences, G\"{o}ttingen 37077, Germany}
\email{agodec@mpinat.mpg.de}

\begin{abstract}
  It was 
  predicted and recently experimentally confirmed that
systems with microscopically reversible 
dynamics in locally \blue{quadratic} potentials warm up
faster than they cool down. This thermal relaxation asymmetry
challenged the local-equilibrium paradigm valid 
near equilibrium. Because the intuition and 
proof 
hinged on the dynamics obeying detailed
balance, 
\blue{it was not clear whether the asymmetry persists}
in systems
with irreversible dynamics. 
\blue{To fill this gap,} 
we here
prove the 
relaxation asymmetry for 
systems driven out of
equilibrium by a general linear drift. The asymmetry persists
due to a
non-trivial isomorphism between driven and reversible
processes. Moreover, rotational motions emerge that, strikingly,
occur in opposite directions during heating and cooling. This
highlights that noisy systems do \emph{not} relax by passing through local
equilibria.
\end{abstract}

\maketitle

According to the laws of thermodynamics, systems in contact
with a thermal environment 
evolve to the temperature of their
surroundings in the process called \emph{thermal relaxation}
\cite{MazurGroot}. Relaxation close to
equilibrium may be explained by linear response theory 
conceptually
based on Onsager's regression hypothesis
\cite{onsager_1,onsager_2,Yokota}. That is, relaxation from a
temperature quench 
is indistinguishable from the
decay of a spontaneous thermal fluctuation at equilibrium
\cite{onsager_1,onsager_2,Yokota}. Analogous results were meanwhile 
formulated also for relaxation near non-equilibrium steady states
\cite{Baiesi_2013,Maes,Polettini}. Beyond the linear regime, however, the regression hypothesis and perturbative arguments fail.   

Important advances have been made in understanding relaxation beyond
the linear regime addressing
\blue{hydrodynamic limits \cite{Bertini2004JSP,Bertini2015RMP}, barrier crossing in driven systems \cite{Maier1993PRE,Bouchet2016AHP}}, memory effects
\cite{Sancho,Igor,Igor_2,Igor_3,Dima_2013,Dima_2019,Lapolla_2019,Lapolla_2020,Haenggi_RMP,Netz},
far-from-equilibrium fluctuation-dissipation theorems
\cite{Dean,Lippiello}, optimal heating/cooling protocols \cite{optimal}, 
anomalous relaxation also known as the \emph{Mpemba effect} \cite{Lu2017MMpemba,Lasanta2017Mpemba,Baity2019MpembaSG,Kumar2020EMpemba,Carollo2021QMpemba,Kumar2022Anomalous,Klich,Oren_2022}
and its isothermal analogue \cite{Deguenther2022EPL}, the 
Kovacs effect
\cite{Kovacs_protein,Militaru2021Kovacs},
and dynamical phase transitions 
\cite{Gaw,Cusp_3,Garrahan,Speck,Minimal,singu_2,Meibohm,Meibohm_2023,Kristian}. Important
advances further include transient
thermodynamic uncertainty relations 
\cite{Pietzonka2017PRE,Dechant_2018,Liu2020PRL,Koyuk2019PRL,Koyuk2020PRL,Dieball2023PRL},
speed limits
\cite{CSL_3,CSL_4,Ito_det,Ito_PRX,Saito},
and analyses of relaxation from the viewpoint of information
geometry \cite{Saito,Ito_PRX,Ibanez}. 

A particularly striking feature of 
relaxation 
was unraveled with the discovery of the asymmetry between
heating and cooling from thermodynamically equidistant temperature
quenches \cite{Lapolla2020PRL}. That is, it was found that systems
with locally quadratic energy landscapes and microscopically
reversible dynamics heat up faster than they cool down.~Later works
expanded on this result 
\cite{VanVu2021PRR,Manikandan2021PRR,Meibohm2021PRE}. The
asymmetry was
recently quantitatively confirmed by experiments \cite{Ibanez}.

The asymmetry emerges because the entropy production within
the system during heating is more efficient than 
heat dissipation
into the environment during cooling \cite{Lapolla2020PRL}. In turn, close to
equilibrium they 
become equivalent and 
symmetry is restored \cite{Lapolla2020PRL,Ibanez}. An even deeper understanding
of the asymmetry was recently achieved by means of ``thermal kinematics''
\cite{Ibanez}. However, both the reasoning and the 
proof 
of the asymmetry \cite{Lapolla2020PRL,Erratum,Ibanez} seem to hinge on the reversibility of the
dynamics. Therefore, the persistence of the 
asymmetry in systems driven 
into
non-equilibrium steady states (NESS) was unexpected. In particular,
a non-conservative
force profoundly changes 
relaxation behavior
\cite{Hwang2005,Kapfer2017PRL,ReyBellet2015,Coghi2021PRE} 
even near stable fixed points \cite{Qian2013JMP} and in systems with
linear drift \cite{Hwang1993}, and may thus \emph{a priori} also break the
asymmetry.   
\begin{figure}[ht!!]
\centering
\includegraphics[width=.48\textwidth]{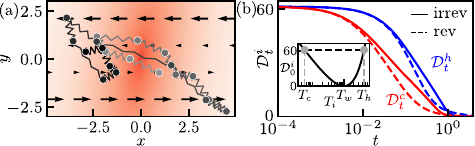}
\caption{(a)~Configuration of a harmonically confined (color gradient)
  Rouse polymer with ${N=20}$ beads in 3d with hydrodynamic interactions
  and internal friction  subject to a shear flow (arrows) in the
  $x$-$y$-plane drawn from the NESS  with covariance $\bSig_{{\rm s},w}$
  (see \cite{Note2} for parameters);
  a projection onto the $x$-$y$-plane is shown.~(b) The corresponding
  free energy difference $\mathcal D^i_t$ in Eq.~\eqref{KL} during
  heating from $T_c$ (red) and cooling from $T_h$ (blue) with (solid lines)
  and without (dashed lines) irreversible shear flow.~The shear 
  changes
  $\mathcal D_t^i$, but the thermal relaxation asymmetry ${\mathcal
    D_t^c<\mathcal D_t^h}$ for $t>0$ remains valid.~Inset: Temperatures $T_i$ before the quench are chosen thermodynamically equidistant, i.e.\ $\mathcal D_0^c=\mathcal D_0^h$.
}\label{Fg1} 
\end{figure}

Here, we investigate 
the speed
and asymmetry of thermal relaxation to an NESS. As a paradigmatic example
we first consider 
a harmonically confined Rouse
polymer with hydrodynamic interactions and internal friction driven by
shear flow (see Fig.~\ref{Fg1}), and demonstrate that heating is
faster than cooling. Next we
provide a systematic analysis of relaxation 
under broken
detailed balance
and explain under which conditions heating and cooling
both become faster.~Finally, we prove that \emph{all} ergodic systems with a
linear drift, including those driven arbitrarily far from equilibrium and
displaying 
rotational motions, heat up faster
than they cool down. In this regime
the notion of a local effective non-equilibrium temperature is
\emph{nominally} impossible.~Our proof, which exploits dual-reversal symmetry,
unravels a non-trivial isomorphism between reversible and driven 
systems.~Finally, we find a new unexpected facet of the relaxation
asymmetry---rotational motions occur 
in \emph{opposite directions} during heating and cooling,
respectively.

\indent \emph{Setup and 
motivating example.---}The relaxation asymmetry was originally
proven for reversible diffusions in locally quadratic
energy landscapes as well as their low-dimensional projections
\cite{Lapolla2020PRL,Erratum}. It states that such systems, when
quenched from thermodynamically equidistant (TED) temperatures $T_h,T_c$ to
an ambient temperature $T_w$ with ${T_c<T_w<T_h}$, heat up
faster than they cool down. In quantitative terms, the generalized
excess free energy in units of $k_{\rm B}T_w$
\cite{Lebowitz1957AP,Mackey1989RMP,Qian2013JMP,Vaikuntanathan2009EPL}
or non-adiabatic entropy production
\cite{Esposito2010PRL,VandenBroeck2010PRE} (i.e.\ the relative entropy
in units of $k_{\rm B}$ \cite{Kullback1951AMS} between the
instantaneous $P_i^w(\x,t)$ and stationary $\ps^w(\x)$ probability
density at $T_w$ with $i=h,c$)
\begin{align}
\!\!\!\mathcal D^{i}_t\!\equiv\!\mathcal D_{\rm KL}[P_i^w(\x,t)||\ps^w(\x)]\!\equiv\!\!\int\!\!\rmd\x
P_i^w(\x,t)\ln\frac{P_i^w(\x,t)}{\ps^w(\x)},\!\!
\label{def DKL} 
\end{align}
is always smaller during heating \cite{Lapolla2020PRL,Erratum}. That is,
$\mathcal{D}^{c}_t<\mathcal{D}^{h}_t$ for all $t>0$ and all TED $T_h$ and $T_c$.      

In a strict sense, the 
asymmetry is to be understood as a statement
about linearized drift around a local minimum in some high-dimensional
energy landscape \cite{Lapolla2020PRL}; counterexamples for diffusion in rugged landscapes \cite{Lapolla2020PRL} and for small
quenches also in sufficiently anharmonic 
wells \cite{Meibohm2021PRE}
are known. The generalization to driven systems therefore
involves a linear drift that, however, does not derive from a
potential and breaks detailed balance. \blue{Our main result is the
  discovery and proof (see last section) of the asymmetry $\mathcal{D}^{c}_t<\mathcal{D}^{h}_t$ in driven systems.}

Consider a  $d$-dimensional system evolving according to the overdamped Langevin equation \cite{Ikeda1981,Gardiner1985}
\begin{align}
  \rmd\x_t&=-\f A\x_t\rmd t+\bsig_i\rmd\f W_t,\label{SDE}
\end{align}
with square drift  and noise-amplitude matrices, $\f A$
and $\bsig_i$, respectively.~In
terms of the 
friction matrix $\bgam$, given by Stokes' law, the positive definite diffusion matrix
reads $\f D_i\equiv\bsig_i\bsig_i^T/2=k_BT_i\bgam^{-1}$ and thus depends
linearly on temperature $T_i$.~The external
force $\f F(\x)$ yields a $T_i$-independent drift 
$-\f A\x=\bgam^{-1}\f F(\x)$, where $\f A$
is generally non-symmetric but confining, i.e.\ the eigenvalues of $\f A$ have
positive real parts.~Thus, $\x_t$
is ergodic but irreversible with zero-mean 
  Gaussian NESS
density $\ps^i(\x)=(2\pi)^{-d/2}\det[\bSig_{{\rm s},i}]^{-1/2}\exp[-\f
  x^T\bSig_{{\rm s},i}^{-1}\f x/2]$ where the covariance $\bSig_{{\rm s},i}$ 
obeys the Lyapunov equation \cite{Note2}
\begin{align}
\f A\bSig_{{\rm s},i}+\bSig_{{\rm s},i}\f A^T=2\mathbf D_i=2k_BT_i\bgam^{-1},\label{Lyapunov ss}
\end{align}  
and thus depends linearly on the temperature $T_i$.~Eq.~\eqref{Lyapunov
  ss} implies for all $T_i$ 
the decomposition into reversible ${-\f A_{\rm rev}\x\equiv\f
D_i\nabla\ln \ps^i(\x)=-\f D_i\bSig_{{\rm s},i}^{-1}\x}$ and irreversible ${{-\f
A_{\rm irr}\x\equiv(-\f A+\f A_{\rm rev})\x}=
{-\balpha_i\bSig_{{\rm s},i}^{-1}\x}}$ drift \cite{Dieball2022PRR}, where
${\balpha_i^T=-\balpha_i}$ is an antisymmetric matrix \footnote{As $\bSig_{{\rm s},i}$ is invertible and
symmetric Eq.~\eqref{Lyapunov ss} 
${\balpha_i=(\f A-\f D_i\bSig_{{\rm s},i}^{-1})\bSig_{{\rm s},i}}={-\bSig_{{\rm s},i}(\f
A^T-\bSig_{{\rm s},i}^{-1}\f D_i)}=-\balpha_i^T$.~In fact $\bSig_{{\rm s},i}^{-1}\x$ and ${\balpha_i\bSig_{{\rm s},i}^{-1}\x}$ are orthogonal since  
their scalar product 
yields an antisymmetric quadratic form $\x^T\bSig_{{\rm s},i}^{-1}{\balpha_i\bSig_{{\rm s},i}^{-1}\x}=0$ \cite{Qian2013JMP}.}.

We focus on temperature
quenches---
instantaneous changes of the environmental temperature
at fixed drift. The thermodynamics of relaxation upon a quench
${T_i\to T_w}$ is
fully specified by 
$\mathcal{D}_t^i$, as the adiabatic entropy
production (housekeeping heat divided by $T_w$)
\cite{VandenBroeck2010PRE} merely embodies the cost of maintaining the NESS
\cite{Seifert2012RPP} and thus need not be considered.~Therefore, 
TED temperatures $T_{h,c}$ correspond to
$\mathcal{D}_0^h=\mathcal{D}_0^c$ and are equal to those of a
reversible system
at the same $T_w$ \cite{Lapolla2020PRL}. 

Since the initial condition is a zero-mean Gaussian with
$\bSig_i^w(0)=\bSig_{{\rm s},i}$,
the
probability density 
is Gaussian for all
times with 
$\bSig_i^w(t)\equiv\E{\x_t\x_t^T}^w_i-\E{\x_t}^w_i\E{\x_t^T}^w_i$ given by \footnote{See Supplemental Material at [\dots].}
\begin{align}
\textcolor{black}{\frac{\rmd}{\rmd t}\bSig_i^w(t)}&\textcolor{black}{\,\,=-\f A\bSig_i^w(t)-\bSig_i^w(t)\f A^T+2\mathbf D_w}\nonumber\\
\Rightarrow\bSig_i^w(t)&=\bSig_{{\rm s},w}+\rme^{-\f A
    t}\left[\bSig_{{\rm s},i}-\bSig_{{\rm s},w}\right]\rme^{-\f A^Tt},
  \label{covariance in time}
\end{align}
where $\langle\cdot\rangle_i^w$ denotes the average over all paths $\x_t$ at temperature $T_w$
evolving from $\ps^i(\x)$. Note that
$\bSig_{{\rm s},i}=T_i\bSig_{{\rm s},w}/T_w$ [see~Eq.~\eqref{Lyapunov ss}]. Introducing $\delta \T_i\equiv T_i/T_w-1$, the generalized excess free energy
reads (see \cite{Note2})
\begin{align}
\mathcal D_t^i=\frac{1}{2}\delta\T_i\,\tr \f X(t)-\frac{1}{2}\ln\det\left[\mathbbm 1+\delta\T_i\,\f
  X(t)\right],
\label{KL}
\end{align}
where we introduced the $d\times d$ matrix
\begin{align}
\f X(t)\equiv \rme^{-\f A t}\bSig_{{\rm s},w}\rme^{-\f A^T t}{\bSig_{{\rm s},w}^{-1}},
\label{early def X} 
\end{align}
which via Eq.~\eqref{KL} fully describes relaxation dynamics.

As a paradigmatic example for such processes we consider a
harmonically confined Rouse polymer
with $N$ beads experiencing hydrodynamic interactions
\cite{Zimm1956JCP,Doi1988} and internal friction
\cite{Cheng2013JCP,Soranno2012,Schuler2016,Daldrop2018} subject to a shear flow, which was investigated
experimentally in
\cite{Smith1999S,Gerashchenko2006PRL,Doyle2000PRL,Perkins1995S,Schroeder2005PRL,Teixeira2005M,Harasim2013PRL,AlexanderKatz2006PRL,Perkins1997S}. 
For a
representative configuration of the NESS ensemble, see
Fig.~\ref{Fg1}a. One may also consider colloidal particles in the
presence of non-conservative optical forces
\cite{Roichman2008PRL}.~The effect of these forces is included
in the $3N\times 3N$ drift matrix $\f A$ and $3N\times 3N$ noise
amplitude $\bsig_i$ \cite{Note2}. Evaluating $\mathcal D_t^i$ for the heating and
cooling processes upon quenches from TED temperatures $T_h$ and $T_c$
we find $\mathcal D_t^c<\mathcal D_t^h$ for all $t>0$.~That is, heating is faster than
cooling (the red line in Fig.~\ref{Fg1}b is
at all times below the blue line). This 
agrees with the
relaxation asymmetry predicted \cite{Lapolla2020PRL} and
experimentally verified \cite{Ibanez} in 
reversible
systems, and provokes the question if this holds 
for any linear driving.
\begin{figure*}[ht!!]
\centering
\includegraphics[width=.95\textwidth]{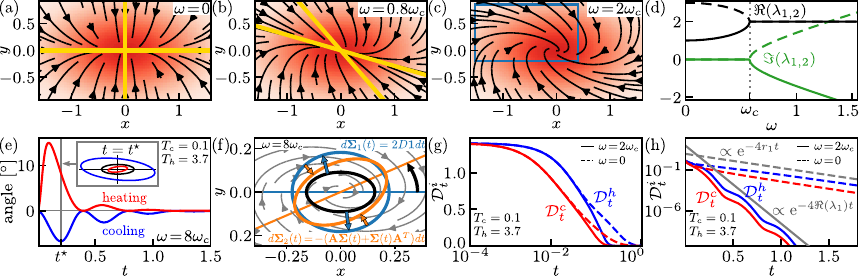}
\caption{\blue{(a-c) Steady-state density $\ps^w(\x)$ (color gradient) 
and streamlines of the drift field $-\f A\x$
for a 2d motion in Eq.~\eqref{SDE} with $\bsig_w=\sqrt{2}\mathbbm 1$ and drift matrix 
$\f A$ with elements $A_{jj}=r_j$ with $r_1=1$, $r_2=3$, 
$A_{jk}=(-1)^{j}\omega r_k$ for ${j,k\in\{1,2\}}$, with $\om$ in 
units of $\om_c\equiv\abs{r_2-r_1}/2\sqrt{r_1r_2}$. 
Real eigendirections (yellow) only exist for $\om\le\om_c$. 
(d)} Real and imaginary parts of
eigenvalues of $\f A$ as a function of $\om$. At $\om=\om_c$
the eigenvalues coincide and eigendirections (yellow lines in b,c)
merge, i.e.\ $\f A$ is not diagonalizable. For $\om>\om_c$
the eigenvalues are complex.
(e) Angle between the covariance matrices
$\bSig_i^w(t)$ and $\bSig_{{\rm s},w}$.~\blue{(f)~Explanation of the
  counter-intuitive 
  opposing
  (effective) rotations at small times during heating from
  $T_c/T_w=0.1$. The change $\rmd\bSig(t)$ in Eq.~\eqref{covariance
    in time} starting from the initial $\bSig_{{\rm s},i}$ 
  (black ellipse) for $\rmd t=0.05$ split into diffusive (yielding the
  blue ellipse) and drift along the grey streamlines (yielding orange ellipse) contributions.}
(g-h) $\mathcal D_t^i$ for heating and cooling 
with and without driving
on logarithmic-linear and linear-logarithmic
scales. The driven system relaxes faster at large $t$
as predicted from the eigenvalues in (e).~Grey lines in (h) show the
limiting relaxation rates for long times, $\rme^{-4r_1t}$ (dashed
line) and $\rme^{-4\Re(\lambda_1)t}$ (solid line).
}
\label{Fg2}
\end{figure*}

\indent \emph{Systematics of breaking detailed balance.---}
We now systematically assess the
influence of non-equilibrium drifts on 
relaxation 
upon a temperature
quench.~As shown above,  \emph{any} linear drift $\f A$ for ${i={c,w,h}}$ 
decomposes as 
\begin{align}
\f A=(\f D_i+\balpha_i)\bSig_{{\rm s},i}^{-1}\quad\rm with\quad \balpha_i^T=-\balpha_i.\label{drift construction}
\end{align}
Thus, by choosing any antisymmetric matrix $\balpha_i$ we alter the NESS
current as well as $\f X(t)$, but neither $\bSig_{{\rm s},i}$ nor $\ps(\x)$. We
can thus directly compare an NESS with the corresponding reversible system
$\balpha_i=\f 0$ with the same steady state. Note that such a direct
comparison is \emph{not} given in the example in Fig.~\ref{Fg1}, since
the shear flow alters $\bSig_{{\rm s},i}$ as it is not of the form
$\balpha_i\bSig_{{\rm s},i}^{-1}$ with $\balpha_i^T=-\balpha_i$ (see
\cite{Note2} for details about the \emph{consistent} comparison of
equilibrium versus nonequilibrium). 

We now consider influence of the non-equilibrium driving.~For linear
drift 
the
relaxation is governed by the eigenvalues of $\f A$
\cite{Metafune_1,Metafune_2}. 
Since $\bSig_{{\rm s},i}$ is, by definition, symmetric with positive
eigenvalues, we can find a matrix $\bbeta=\bbeta^T$ such that
$\bbeta^2\equiv\bSig_{{\rm s},i}^{-1}$ \footnote{From the orthogonal
diagonalization $\f O \bSig_{{\rm s},i}^{-1} \f O^{T}={\rm
  diag}(s_j)$ we define $\bbeta\equiv \f O^T {\rm diag}\sqrt{s_j}\f O$}.~Thus, the matrix $\bbeta\f
D_i\bSig_{{\rm s},i}^{-1}\bbeta^{-1}=\bbeta\f
D_i\bbeta=\bbeta\bsig_i(\bbeta\bsig_i)^T/2$ is symmetric which alongside
$\det(\bbeta\bsig_i)\ne 0$ implies that $\f D_i\bSig_{{\rm s},i}^{-1}$ is
diagonalizable with positive eigenvalues 
\footnote{Any matrix of the form $\f
M\f M^T$ is symmetric, and therefore diagonalizable, with real
non-negative eigenvalues, since $\f M\f M^T\f v=\lambda\f v$ implies
$\lambda=\f v^T\f M\f M^T\f v/\f v^T\f v=(\f M^T\f v)^T\f M^T\f v/\f
v^T\f v\ge 0$}. Therefore, in the absence of driving $\f A=\f
D_i\bSig_{{\rm s},i}^{-1}$ expectedly has strictly positive eigenvalues
reflecting a monotonous relaxation to equilibrium.

Once we include driving $\balpha_w\ne 0$ in the steady-state-preserving form
Eq.~\eqref{drift construction}, the spectrum may or may not become
complex depending on the detailed form of $\balpha_w$, see
e.g.\ \blue{Fig.~\ref{Fg2}a-d}.\ Complex eigenvectors imply that eigendirections where the drift
points ``straight'' towards $\f 0$ cease to exist, see
\blue{Fig.~\ref{Fg2}a-c}.~This happens already
at arbitrarily small driving if level sets of $\ps(\x)$ are
(hyper)spherical. 
If some eigenvalues are on the threshold
of becoming complex (branching point $\omega_c$ in \blue{Fig.~\ref{Fg2}d}), $\f A$ may become non-diagonalizable. In terms of
the minimal 2d 
example in Fig.~\ref{Fg2} we have that $\f
A$ 
is non-diagonalizable when
$\om=\pm\om_c$ (see \blue{Fig.~\ref{Fg2}d}). 

An interesting consequence of driving is that the different dimensions
no longer decouple as they do under detailed balance (see
\blue{Fig.~\ref{Fg2}a}).~This means that the $d$-dimensional Langevin equation
\eqref{SDE} cannot be decomposed into 1d 
equations and
that rotational dynamics may emerge.
In the particular case of
temperature quenches 
we find that driving causes a time-dependent rotation of the level sets of
$P_i^w(\x,t)$, see \blue{Fig.~\ref{Fg2}e.~In agreement with the
opposite sings of $T_i-T_w$ in Eq.~\eqref{covariance in time}, these
rotations occur in
opposite directions during heating and cooling, which} is a striking new feature of
the relaxation asymmetry.~The asymmetry implies that thermal relaxation must
\emph{not} be understood as passing through local
equilibria at intermediate (effective) temperatures \cite{MazurGroot}, since this would
imply a symmetric relaxation independent of the sign of
the temperature quench.
Moreover, the rotation in opposite directions emphasizes
that heating and cooling here evolve along very distinct pathways in
the space of probability distributions (see also \cite{Ibanez}).\\
\indent\blue{While the initial rotation during cooling follows the direction
  of driving,  most surprisingly the effective rotations during
  heating initially oppose the direction of the driving (see
  Fig.~\ref{Fg2}e). This effect can be traced to the interplay of
  (``Trotterized'' \cite{Trotter}) diffusion and drift during individual small time increments, 
  see Fig.~\ref{Fg2}f. During heating for an increment  $\rmd t$ 
  diffusion alone 
  propagates the black to the more circular blue
  ellipse. The subsequent drift along the elliptical streamlines
  propagates this blue ellipse to the orange ellipse that is, however,
  effectively rotated in the direction opposite to the drift (for
  further details see \cite{Note2}).\\
  \indent \emph{Accelerated relaxation.---}Before proving the relaxation
asymmetry we discuss the acceleration of relaxation via driving
\cite{Hwang1993,Hwang1993,Kapfer2017PRL,ReyBellet2015,Coghi2021PRE}. We therefore} focus on the real part of the eigenvalues which determines the
relaxation time-scales. Upon a change of basis we find 
 $\widetilde{\f A}\equiv\bbeta\f A\bbeta^{-1}=\bbeta\f
D_i\bbeta+\bbeta\balpha_i\bbeta$ where
$(\bbeta\balpha_i\bbeta)^T=-\bbeta\balpha_i\bbeta$.
Then, for any complex eigenvalue $\lambda$ of $\widetilde{\f A}$ with
eigenvector $\f v\ne 0$ we may write $2\Re(\lambda)\f v^\dag\f
v=(\lambda+\lambda^\dag)\f v^\dag\f v=\f v^\dag(\widetilde{\f A}+\widetilde
{\f A}^\dag)\f v=2\f v^\dag\bbeta\f D_i\bbeta\f v$, where $\dag$ denotes the
Hermitian adjoint. Decomposing $\f v$, $\f v^\dag$
in the orthonormal eigenbasis of $\bbeta\f D_i\bbeta$ with
eigenvalues $0<\mu_1\le\dots\le\mu_d$, we have with $c_j\in\C$
\begin{align}
\Re(\lambda)=\frac{\f v^\dag \bbeta\f D_i\bbeta\f v}{\f v^\dag\f v}=\frac{\sum_{j=1}^d c_j^\dag c_j\mu_j}{\sum_{j=1}^d c_j^\dag c_j}\in[\mu_1,\mu_d].\label{proof faster relaxation}
\end{align}
This means that the real parts of the eigenvalues in the presence of
driving remain not only positive, as required for the existence of
a steady state, but even remain in the interval $[\mu_1,\mu_d]$. 
\blue{Thus, }Eq.~\eqref{proof faster relaxation} states that the smallest real part
of eigenvalues of $\f A$ under driving obeys
$\Re(\lambda_1)\ge\mu_1$.~Note that $\Re(\lambda_1)$
typically \footnote{Unless the initial distribution has only a negligible projection
onto the slowest modes.} sets the slowest relaxation rate
\cite{Metafune_1,Metafune_2}. Since $\Re(\lambda_1)$ increases (or
does not
decrease) upon driving, the latter typically enhances relaxation on
long time scales\blue{, as already shown
  in \cite{Hwang1993}}.

Driving also affects the adiabatic entropy production.~This effect, however, scales trivially, as  the adiabatic entropy production increases
with increasing $\balpha_i$ according to
$\balpha_i^T\f D^{-1}_i\balpha_i$ \cite{Note2}.~Hence, there is \emph{no direct
connection} between faster relaxation and steady-state dissipation,
as the influence of driving on the eigenvalues is specific.~For
example, the acceleration in $d=2$ saturates [see $\Re(\lambda_1)$ in
\blue{Fig.~\ref{Fg2}d}].~More drastically, 
multiplying $\balpha_i$ by a factor larger than $1$ in ${d=3}$ may
decrease $\Re(\lambda_1)$ \cite{Hwang1993}.

We see from Eq.~\eqref{early def X} that ${\f
X(t)\sim\rme^{-2\Re(\lambda_1)t}}$ for long times and therefore  ${\mathcal D_t^i\sim\rme^{-4\Re(\lambda_1)t}}$ (see
\cite{Note2} and Fig.~\ref{Fg2}g-h). The statement ``accelerated
relaxation'', 
$\Re(\lambda_1)\ge\mu_1$, means that
both, heating and cooling will 
at long times be
faster.~In general the \emph{difference} between heating and cooling upon
driving can become 
larger or smaller than for reversible dynamics with the same
$\bSig_{{\rm s},i}$, but as we now prove heating is always faster
than cooling.

\begin{figure}[ht!!]
\begin{center}
\includegraphics[width=.45\textwidth]{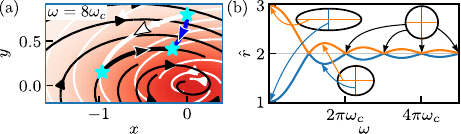}
\caption{(a) Illustration of Eq.~\eqref{dual reversal form of
    X}:~Streamplot of the drift field $-\f A\x$ (black) as in blue frame in
  \blue{Fig.~\ref{Fg2}c}, and inverted
  drift field $-\f A_{-\balpha}\x$ (white). The white line depicts $\rme^{-\f
  A_{-\balpha}\tau}\x_0$ for $\tau\in[0,t]$, the black line is
  $\rme^{-\f A\tau}\rme^{-\f A_{-\balpha}t}\x_0$, and the blue line
  shows $\f X(\tau)\x_0$. (b) Effective stiffness
  $\hat r_j(\om)\equiv-\ln(x_j^t)/2t$ at ${t=1}$ as a function of driving
  $\om$ (see \cite{Note2}). For large driving the directions mix, such that the system
  effectively approaches a circular parabola with stiffness
  $(r_1+r_2)/2$, which is the real part of eigenvalues in \blue{Fig.~\ref{Fg2}d}.
}\label{Fg3}
\end{center}
\end{figure}

\indent \emph{Proof of relaxation asymmetry in driven systems.---}
We now prove the relaxation asymmetry for the dynamics in
Eq.~\eqref{SDE}, i.e.\ $\Delta\mathcal D_t\equiv\mathcal
D_t^h-\mathcal D_t^c>0$ for all $t>0$.
By Eq.~\eqref{early def X} 
\begin{align}
\!\!\!\Delta\mathcal D_t=\frac{\delta \T_h-\delta \T_c}{2}\,\tr \f
X(t)-\frac{1}{2}\ln\frac{\det\left[\mathbbm 1+\delta\T_h\,\f X(t)\right]}{\det\left[
  \mathbbm 1+\delta\T_c\,\f
    X(t)
    \right]}.
\label{delta DKL}
\end{align}
To prove the
asymmetry we must understand the properties of $\f X(t)$, which
is $T_i$-independent. Using the
steady-state Lyapunov equation \eqref{Lyapunov ss} we can rewrite $\f
X(t)$ as
\begin{align}
\f X(t)=\rme^{-\f At}\rme^{-\f A_{-\balpha}t},\label{dual reversal form of X} 
\end{align}
where $\f A_{-\balpha}\equiv(\f D_w-\balpha_w)\bSig_{{\rm s},w}^{-1}$ is the driving-reversed version of $\f A$ as in Eq.~\eqref{drift construction}. This form is reminiscent of the dual-reversal symmetry \cite{Hatano2001PRL,Dechant2021PRR,Dieball2022PRL,Dieball2022PRR} stating that time-reversal in non-equilibrium steady states requires concurrent current reversal. Eq.~\eqref{dual reversal form of X} is illustrated in Fig.~\ref{Fg3}a. 
The proof again requires to change the basis via $\bbeta$ as
\begin{align}
\widetilde{\f X}(t)\equiv \bbeta\f X(t)\bbeta^{-1}=\rme^{-\widetilde{\f A}t}\left(\rme^{-\widetilde{\f A}t}\right)^T,
\label{MMT}
\end{align}
where we used $\bbeta{\f A}_{-\alpha}\bbeta^{-1}=\widetilde{\f A}^T$ and
$\rme^{-\widetilde{\f A}^Tt}=(\rme^{-\widetilde{\f A}t})^T$. Thus,
$\widetilde{\f X}(t)$ is symmetric and hence diagonalizable with real
eigenvalues.~Since, $\det \rme^{-\widetilde{\f A}t}=\rme^{-\tr
  \widetilde{\f A}t}$, we have $\det \widetilde{\f
  X}(t)=\rme^{-2\tr \widetilde{\f A}t}\ne 0$. Therefore, $\widetilde{\f X}(t)$
and thus ${\f X}(t)$ have positive eigenvalues $x^t_j>0$, ${j=1,\dots,d}$ \cite{Note4}.~Although $\f A$ may have complex eigenvalues or even be
non-diagonalizable and $\exp(-\f At)$ may be rotational (see
\blue{Figs.~\ref{Fg2}c} and \ref{Fg3}a), $\f X(t)$ has a real eigensystem
since consecutive rotations in forward and
current-reversed directions effectively cancel rotations, see Eq.~\eqref{dual reversal form of X} and Fig.~\ref{Fg3}a.

Using the eigenvalues $x_j^t>0$ we
rewrite Eq.~\eqref{delta DKL} as 
\begin{align}
\Delta\mathcal D_t
=&\sum_{j=1}^d\left(
\frac{\delta\T_h-\delta \T_c}{2}x^t_j-\frac{1}{2}\ln\left[\frac{1+\delta\T_hx^t_j}{1+\delta\T_cx_j^t}\right]\right).\label{delta DKL eigenvalues}
\end{align}
If all  ${x^t_j\in(0,1)}$, 
the proof for reversible systems
\cite{Lapolla2020PRL,Erratum} asserts that $\Delta\mathcal D_t>0$. It
therefore suffices to show
that $x^t_j<1$ for all $j$, which is equivalent to $\norm{\f X(t)}<1$, where $\norm{\f M}\equiv\sup_{\f
  v\in\R^d\setminus\f 0}\norm{\f M\f v}_2/\norm{\f v}_2$ and $\norm{\f
  v}_2=\sqrt{\f v^T\f v}$ are the matrix and Euclidean norm,
respectively.~Eq.~\eqref{dual reversal form of X} does not help in
showing this \footnote{Eq.~\eqref{dual reversal form of X} suffices
only at
equilibrium $\f A=\f A_{-\balpha}=\f D_w\bSig_{{\rm s},w}^{-1}$ where $\f X(t)=\exp(-2\f
D_w\bSig_{{\rm s},w}^{-1} t)$ decomposes into $x^t_j=\exp(-2\mu_jt)<1$.}; although eigenvalues  of $\f A$ have positive real
parts [see Eq.~\eqref{proof faster relaxation}], it may be that
$\norm{\rme^{-\f A_{\pm\balpha}t}}>1$
  (e.g. the distance to $\f 0$ in Fig.~\ref{Fg3}a
increases along the white line).~This is possible because the
eigenvectors of $\f A$ are not orthogonal. 

We thus change the basis as in Eq.~\eqref{MMT} and use
the log-norm inequality $\norm{\exp(\f Mt)}\le\exp[\mu(\f M)t]$
\cite{Dahlquist1958} with log norm \blue{$\mu(\f M)\equiv\lim_{h\to0^+}h^{-1}\left({\norm{\mathbbm 1+h\f M}-1}\right)$ yielding }$\mu(-\widetilde{\f A})\equiv\mu(-\bbeta\f A\bbeta^{-1})=\mu(-\bbeta\f D_w\bbeta)=-\mu_1$
determined by the symmetric part $(\widetilde{\f A}+\widetilde{\f
  A}^T)/2=\bbeta\f D_w\bbeta$ \cite{Note2}.~This basis is appropriate because 
$\bbeta\balpha_w\bbeta$ in $\widetilde{\f A}$ (unlike
$\balpha_w\bSig_{{\rm s},w}^{-1}$ in $\f A$) has no symmetric part,
i.e.\ 
the driving only affects the rotational part. 
The log-norm inequality thus implies $\norm{\exp(-\widetilde{\f A}t)}\le\exp[\mu(-\widetilde{\f A})t]=\exp(-\mu_1t)$ and similarly $\norm{\exp(-\widetilde{\f A}^T t)}\le\exp(-\mu_1t)$, and by the submultiplicative property of the matrix norm we obtain from Eq.~\eqref{MMT} 
\begin{align}
\norm{\widetilde{\f X}(t)}\le\Big|\Big|{\rme^{-\widetilde{\f A}t}}\Big|\Big|\ \Big|\Big|\left(\rme^{-\widetilde{\f A}t}\right)^T\Big|\Big|\le\rme^{-2\mu_1t}<1.
\end{align}
Since ${\norm{\widetilde{\f X}(t)}=\norm{\f X (t)}}$ this implies $x^t_j<1$ and with Eq.~\eqref{delta DKL eigenvalues}
completes the proof of $\Delta\mathcal D_t>0$ for all $t>0$.

The proof provides important insight into
the thermodynamics of the asymmetry 
in reversible versus
driven systems. Namely, $\Delta \mathcal D_t$ in Eq.~\eqref{delta DKL eigenvalues} for a driven
system at any $t$ is equal to that of any reversible system with
drift matrix $\hat{\f A}$ having eigenvalues
$\hat{\mu}_i$ satisfying
$\rme^{-2\hat{\mu}_jt}=x^t_j$. Therefore,
at each $t$ the relaxation asymmetry of a driven system is isomorphic to
that of an equilibrium system with different geometry (see Fig.~\ref{Fg3}b for
effective stiffness axes of the 2-dimensional
parabolic potential), which implies the persistence of the asymmetry.~This provokes intriguing questions about the
existence of the
asymmetry in the presence of time-dependent driving. 

\indent \emph{Conclusion.---}We have proven that overdamped
ergodic systems driven by linear drift, conservative or not, for any
pair of thermodynamically equidistant temperature quenches warm up
faster than they cool down. The 
relaxation asymmetry \cite{Lapolla2020PRL}, which was recently
confirmed experimentally \cite{Ibanez}, therefore persists in driven
systems. As the original proof hinged on
microscopic reversibility, this finding is surprising and is explained
by a non-trivial isomorphism between driven and reversible
processes. In the presence of driving a striking new feature of
the relaxation asymmetry appears: rotational dynamics emerge with
opposite directions during heating and cooling, respectively. This
further highlights that small, noisy systems do \emph{not} relax by passing through local
equilibria 
\cite{MazurGroot}.~Moreover, rotations in opposing directions
emphasize that heating and cooling evolve along fundamentally distinct
pathways \cite{Ibanez}.~An analysis with the framework of ``thermal
kinematics'' \cite{Ibanez} will bring even deeper insight.
Our results motivate further studies on the existence of the relaxation
asymmetry in temporally driven systems
\cite{blickle2012realization,martinez2015adiabatic,martinez2016brownian,krishnamurthy2016micrometre,Koyuk2020PRL,rademacher2022nonequilibrium},
systems with non-linear drift
\cite{Lu2017MMpemba,Felix_2019,Baity2019MpembaSG,Kumar2022Anomalous,Kumar2020EMpemba}, 
and in the presence of inertial effects \cite{Militaru2021Kovacs}.

\emph{Acknowledgments.---}Financial support from Studienstiftung des Deutschen Volkes (to C.\ D.) and the German Research Foundation (DFG) through the Emmy Noether Program GO 2762/1-2 (to A.\ G.) is gratefully acknowledged.

\let\oldaddcontentsline\addcontentsline
\renewcommand{\addcontentsline}[3]{}
\bibliographystyle{apsrev4-2}
\bibliography{bib_ellipse.bib}
\let\addcontentsline\oldaddcontentsline

\clearpage
\newpage
\onecolumngrid
\renewcommand{\thefigure}{S\arabic{figure}}
\renewcommand{\theequation}{S\arabic{equation}}
\setcounter{equation}{0}
\setcounter{figure}{0}
\setcounter{page}{1}
\setcounter{section}{0}

\begin{center}\textbf{Supplemental Material for:\\ Asymmetric Thermal Relaxation in Driven Systems: Rotations go Opposite Ways}\\[0.2cm]
Cai Dieball$^1$, Gerrit Wellecke$^{1,2}$ and Alja\v{z} Godec$^1$\\[0.15cm]
\emph{$^1$Mathematical bioPhysics Group, Max Planck Institute for Multidisciplinary Sciences, 37077 G\"{o}ttingen, Germany}\\
\emph{$^2$Present address: Theory of Biological Fluids, Max Planck Institute for Dynamics and Self-Organization, G\"{o}ttingen 37077, Germany}
\\[0.6cm]\end{center}
\begin{quotation}
In this Supplementary Material we provide further details on
  model examples, arguments, and calculations presented in the
  Letter. Besides several technical details, we give the equations and
  parameters describing the the Rouse chain in confined shear flow and derive and solve equations for the covariance. \blue{We extensively elaborate on the rotations in different directions, and address} the consistent comparison of equilibrium and non-equilibrium steady states. We conclude
with a discussion of the log-norm inequality.
\end{quotation}
\hypersetup{allcolors=black}\tableofcontents\hypersetup{allcolors=mylinkcolor}

\section{Rouse polymer with hydrodynamic interactions and internal friction in confined shear flow}\label{Model}
In Fig.~1 in the Letter we consider the motivating example of a
polymer chain with $N=20$ beads in $3d$ space represented by the Rouse
model with internal friction and
hydrodynamic interactions in shear flow. That is, we assume that the
beads are connected by harmonic springs with zero rest length [Eq.~\eqref{Rouse first equation}]
and
additionally interact via hydrodynamic interactions [Eq.~\eqref{HI}] and experience
internal friction [Eq.~\eqref{with-IF}]. The chain is confined in a parabolic potential and
is subject to a shear flow [Eqs.~\eqref{all_Rouse}-\eqref{Rouse last
  equation}]. We now describe the interactions and evolution equations
individually, with increasing complexity. 

In the classical Rouse model (i.e.\ without hydrodynamic interactions,
internal friction, confinement and shear), the time-dependent position
of the beads $\x_t$ is described by the $3N$ dimensional Langevin
equation (denoting the spring stiffness by $\kappa$ and solvent friction by $\gamma$)
\begin{align}
\gamma\rmd\x_t=-\kappa\f k\x_t\rmd t+\sqrt{2D_i}\rmd\f W_t,\label{Rouse first equation} 
\end{align}
where the connectivity matrix $\f k$ is a $3N\times 3N$ matrix that reads ($\mathbbm{1}_3$ is the 3d unit matrix and all terms not shown are $0$)
\begin{align}
\f k=\begin{bmatrix}\mathbbm 1_3&-\mathbbm 1_3&&&&&\\-\mathbbm 1_3&2\mathbbm 1_3&-\mathbbm 1_3&&&0&\\&-\mathbbm 1_3&2\mathbbm 1_3&-\mathbbm 1_3&&&\\&&&\dots&\dots&&\\&&&&\dots&\dots&\\&0&&&-\mathbbm 1_3&2\mathbbm 1_3&-\mathbbm 1_3\\&&&&&-\mathbbm 1_3&\mathbbm 1_3\end{bmatrix}.
\end{align}
Note that the temperature dependence is contained in the diffusion constant $D_i\propto T_i$ as described in the Letter. Following Ref.\ \cite{SM_Cheng2013JCP}, we introduce hydrodynamic interactions in the pre-averaging approximation via the $3N\times 3N$ matrix $\f H$ with $3\times 3$ entries for $m,n=1,\dots,N$
\begin{align}
H_{mn}=\gamma^{-1}\mathbbm{1}_3\times\begin{cases}1\qquad\qquad\qquad\ \ \:{\rm
  if\ }n=m\\\frac{1}{2\pi^2\gamma\abs{n-m}^{1/2}}\qquad {\rm if\ }n\ne
m\end{cases},
\label{HI}
\end{align}
and also include internal friction with friction parameter $\xi_{\rm
  IF}$, such that the equation of motion of the chain becomes
\begin{align}
\rmd\x_t=\f H\left(-\kappa\f k\x_t\rmd t-\xi_{\rm IF}\f
k\rmd\x_t+\sqrt{2D_i}\rmd\f W_t\right).
\label{with-IF}
\end{align}
We now introduce a spatial confinement via a $3d$ harmonic potential
centered at $\x=\f 0$ and subject the confined chain to a shear flow
in the $x$-$y$-plane. To account for the confinement, we consider the three-dimensional matrix ($r_i>0$)
\begin{align}
\f C=\f R(\theta)\begin{bmatrix}r_x&0&0\\0&r_y&0\\0&0&r_z\end{bmatrix}\f R(\theta)^T,
\label{Rouse C}
\end{align}
and for the shear flow in the $x$-$y$-plane
\begin{align}
\f S=\begin{bmatrix}0&\omega&0\\0&0&0\\0&0&0\end{bmatrix}.
\label{Rouse S}
\end{align}
\blue{Note that by introducing the shear flow $\f S$ to the system, the microscopic reversibility is lost, i.e., we have an example that is genuinely driven out of equilibrium.}

To avoid the special case where shear and confinement are orthogonal,
we introduce the rotation matrix $\f R(\theta)$ to rotate the
confinement by an angle $\theta$ within the $x$-$y-$plane,
\begin{align}
\f R(\theta)=\begin{bmatrix}\cos(\theta)&-\sin(\theta)&0\\\sin(\theta)&\cos(\theta)&0\\0&0&1\end{bmatrix}.\label{Rouse rotation} 
\end{align}
Now consider a $3N$ dimensional matrix $\f M$ with $\f R(\theta)\f C\f
R^T(\theta)+\f S$ on the $N$ diagonal $3\times3$ blocks to impose
the confinement and shear on each bead. Then the
equation of motion for the confined polymer in shear flow reads
\begin{align}
\rmd\x_t=\f H\left(-(\kappa\f k+\f M)\x_t\rmd t-\xi_{\rm IF}\f
k\rmd\x_t+\sqrt{2D_i}\rmd\f W_t\right).
\label{all_Rouse}
\end{align}
This may now be rewritten as
\begin{align}
\left(\f H^{-1}+\xi_{\rm IF}\f k\right)\rmd\x_t&=-(\kappa\f k+\f M)\x_t\rmd t+\sqrt{2D_i}\rmd\f W_t,\nonumber\\
\rmd\x_t&=-\left(\f H^{-1}+\xi_{\rm IF}\f k\right)^{-1}(\kappa\f k+\f M)\x_t\rmd t
+\sqrt{2D_i}\left(\f H^{-1}+\xi_{\rm IF}\f k\right)^{-1}\rmd\f
W_t.\label{Rouse last equation}
\end{align}
Upon identifying
\begin{equation}
\f A = \left(\f H^{-1}+\xi_{\rm IF}\f k\right)^{-1}(\kappa\f k+\f
M),\quad \bsig_i=\sqrt{2D_i}\left(\f H^{-1}+\xi_{\rm IF}\f k\right)^{-1}
\end{equation}  
we arrive at the desired form $\rmd\x_t=-\f A\x_t\rmd t+\bsig_i\rmd\f W_t$ as described in the Letter.

The parameters used in Fig.~1 are $N=20$, $r_x=0$, $r_y=0.1$, $r_z=1$, $\theta=-10^\circ$, $\gamma=1$, $\xi_{\rm IF}=1$, $D_w=1$, $\kappa=1$, $T_c=0.05T_w$, $T_h=4.56T_w$. 
In Fig.~1a we choose $\om=3$ while in Fig.~1b we use $\om=20$ to emphasize the differences between the curves. 
\blue{For the chosen
  parameters, several eigenvalues of the matrix $\f A$ become complex,
  thus confirming that the Rouse model in the shear flow is an
  irreversible process.}

\section{Lyapunov equation and time-dependent covariance}
Here we derive Eqs.~(3) and (4) in the Letter. We consider dynamics
governed by $\rmd\x_t=-\f A\x_t\rmd t+\bsig\rmd\f W_t$ as in Eq.~(2) in the Letter. Taking the mean value gives $\frac{\rmd}{\rmd t}\E{\x_t}=-\f A\E{\x_t}$ which implies $\E{\x_t}=\rme^{-\f A t}\E{\x_0}$. In the Letter we only consider initial conditions with $\E{\x_0}=\f 0$ such that for all times $\E{\x_t}=0$.
The covariance
$\bSig(t)\equiv\E{\x_t\x_t^T}-\E{\x_t}\E{\x_t^T}=\E{\x_t\x_t^T}$ is
always symmetric $\bSig(t)^T=\bSig(t)$ with strictly positive
eigenvalues. Using It\^o's Lemma \cite{SM_Ikeda1981,SM_Gardiner1985} we see that $\bSig(t)$ obeys the differential Lyapunov equation
\begin{align}
\frac{\rmd}{\rmd t}\bSig(t)&=\E{\rmd\x_t\x_t^T}+\E{\x_t\rmd\x_t^T}+\E{\rmd\x_t\rmd\x_t^T}\nonumber\\
&=-\f A\E{\x_t\x_t^T}-\E{\x_t\x_t^T}\f A^T+\bsig\bsig^T\nonumber\\
&=-\f A\bSig(t)-\bSig(t)\f A^T+2\mathbf D.\label{SM Lyapunov differential}
\end{align}
In the steady state (i.e.\ for $\f A$ originating from a confining
potential and $t\to\infty$) this approaches the steady-state
covariance $\bSig_{\rm s}$ obeying the algebraic (i.e.\ non-differential) Lyapunov equation (Eq.~(3) in the Letter, see also Ref.~\cite{SM_Gardiner1985})
\begin{align}
\f A\bSig_{\rm s}+\bSig_{\rm s}\f A^T=2\mathbf D.\label{SM Lyapunov ss}
\end{align}
Given the solution $\bSig_{\rm s}$ of Eq.~\eqref{SM Lyapunov ss}, the solution for Eq.~\eqref{SM Lyapunov differential} for an initial condition with covariance $\bSig(0)$ is obtained as
\begin{align}
\bSig(t)=\bSig_{\rm s}+\rme^{-\f A t}\left[\bSig(0)-\bSig_{\rm s}\right]\rme^{-\f A^Tt}.\label{SM Lyapunov solution} 
\end{align}
This is proven by taking the derivative of the ansatz,
\begin{align}
&\frac{\rmd}{\rmd t}\left(\bSig_{\rm s}+\rme^{-\f A t}\left[\bSig(0)-\bSig_{\rm s}\right]\rme^{-\f A^Tt}\right)\nonumber\\
&=-\f A\left(\rme^{-\f A t}\left[\bSig(0)-\bSig_{\rm s}\right]\rme^{-\f A^Tt}\right)-\left(\rme^{-\f A t}\left[\bSig(0)-\bSig_{\rm s}\right]\rme^{-\f A^Tt}\right)\f A^T\nonumber\\
&=-\f A\bSig(t)+\f A\bSig_{\rm s}-\bSig(t)\f A^T+\bSig_{\rm s}\f A^T\nonumber\\
&\overset{\eqref{SM Lyapunov ss}}=-\f A\bSig(t)-\bSig(t)\f A^T+2\mathbf D.
\end{align}
Choosing $\bSig(0)=\bSig_{{\rm s},i}=T_i\bSig_{{\rm s},w}/T_w$ yields Eq.~(4) in the Letter.

\section{Generalized excess free energy during heating and cooling}

The Kullback-Leibler divergence [Eq.~(1) in the Letter] can be computed for two $d$-dimensional Gaussian densities $P_{1,2}$ with mean zero as $2D_{\rm KL}\left(P_1||P_2\right)=-\ln\left(\det\left[\bSig_1\bSig_2^{-1}\right]\right)+\tr\left(\bSig_1\bSig_2^{-1}-\mathbbm 1\right)$ where $\mathbbm 1$ is the $d$-dimensional unit matrix. Using Eq.~(4) in the Letter (i.e.\ $\bSig_i^w(t)=\bSig_{{\rm s},w}+\rme^{-\f A t}\left[\bSig_{{\rm s},i}-\bSig_{{\rm s},w}\right]\rme^{-\f A^Tt}$) with the notations $\f X(t)\equiv \rme^{-\f A t}\bSig_{{\rm s},w}\rme^{-\f A^T t}{\bSig_{{\rm s},w}^{-1}}$ and $\delta \T_i\equiv T_i/T_w-1$ we obtain Eq.~(5) in the Letter, i.e.
\begin{align}
\mathcal D_t^i=\frac{1}{2}\delta\T_i\,\tr \f X(t)-\frac{1}{2}\ln\det\left[\mathbbm 1+\delta\T_i\,\f X(t)\right].
\end{align}

\blue{
\section{Effective rotations opposing the direction of drift}
In the Letter (see Fig.~2e-f) we state that, and briefly explain why,
rotations of the probability density function (quantified via
covariance ellipses) during heating emerge in opposite directions.
Mathematically opposing rotations can be seen from Eq.~(4) in the Letter,
$\bSig_i^w(t)=\bSig_{{\rm s},w}+\rme^{-\f A t}\left[\bSig_{{\rm
      s},i}-\bSig_{{\rm s},w}\right]\rme^{-\f A^Tt}$, where
$\bSig_{{\rm s},i}-\bSig_{{\rm s},w}=(T_i/T_w-1)\bSig_{{\rm s},w}$ has
opposing signs for $T_c<T_w$ and $T_h>T_w$.  However, the physical or
phenomenological understanding is more challenging.  The most
surprising aspect is that rotational motions occur in directions that 
oppose the rotational driving (e.g.,\ during heating). Since there are
no rotational motions in the absence of driving, clockwise
rotational driving can in fact lead to counterclockwise (effective)
rotations.  

First, note that the phenomenon has to be understood on the level of
probability density functions and  \emph{not} on the level of individual
particles' trajectories. In Fig.~\ref{FgSM1}a,b we show that the cloud of
particles' positions (representing the probability density) effectively rotates
in the counterclockwise direction while the individual particles on average
follow the rotational drift in the clockwise direction, see
Fig.~\ref{FgSM1}c. 
\begin{figure}[ht!!]
\centering
\includegraphics[width=1.\textwidth]{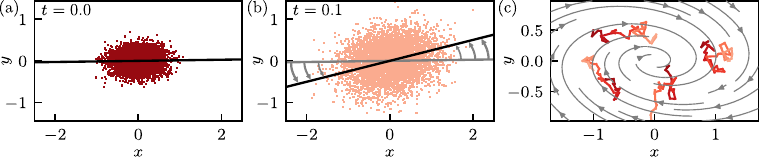}
\caption{\blue{(a,b) Simulation of 5000 particles' trajectories
    evolving according to the two-dimensional overdamped Langevin equation $\rmd\x_t=-\f A\x_t\rmd t+\sqrt{2}\rmd\f W_t$  with $\f A=\begin{bmatrix}
r_1& -r_2\om \\ -r_1\om &r_2\end{bmatrix}$ with $r_1=1$, $r_2=3$,
$\om=8\om_c$ and time-step $\rmd t=0.001$ starting from an initial
condition corresponding to $T_c=0.1$ (in units of $T_w$). (c)
Simulated trajectories of $4$ particles in time $0.0$ (dark) to $0.1$
(bright). Grey streamlines shows $\f A(x,y)^T$, i.e., the direction
that particles' trajectories follow on average.}}
\label{FgSM1} 
\end{figure}

The emergence of this counterintuitive opposing rotation is explained in Fig.~2f in
the Letter. To repeat this, during a Trotterized time-increment the
diffusion propagates the initial covariance
ellipse to a more circular (less eccentric) one. Next,  note that the rotational drift
is \emph{not} a perfect circulation, but instead driving along elliptical
contour-lines \emph{plus} the driving into the center due to the 
confining (conservative) potential. This clockwise elliptical rotational driving
applied to the ellipse (previously ``rounded'' during the diffusion
Trotter-increment)  leads to the counter-clockwise rotation directly by following the streamlines of
the drift (see Fig.~2f in the Letter). 

To elaborate on these rotations consider Fig.~\ref{FgSM2}. Ellipses in
Fig.~\ref{FgSM2} and those shown below are the covariance ellipses, while
ellipses in Fig.~\ref{FgSM1} and Fig.~2f in the Letter correspond to
standard-deviation ellipses (i.e.,\ square roots of covariance
ellipses). In Fig.~\ref{FgSM2}a we recall the opposite rotation during
heating and cooling. As in Fig.~\ref{FgSM1} we then focus on the
heating, where the initial rotation is in the counterintuitive
direction; see Fig.~\ref{FgSM2}b. To illustrate the explanation given in
Fig.~2f in the Letter, we show that this rotation similarly emerges if
we start in a circular initial condition (see Fig.~\ref{FgSM2}c).

If we instead consider a circular driving with circular steady-state
density (see Fig.~\ref{FgSM2}d) all rotations emerge in the
(intuitive) clockwise direction, which shows that the elliptical
(i.e.\ non-circular)
component of the circular driving is a key factor in this
phenomenon. 
\begin{figure}[ht!!]
\centering
\includegraphics[width=1.\textwidth]{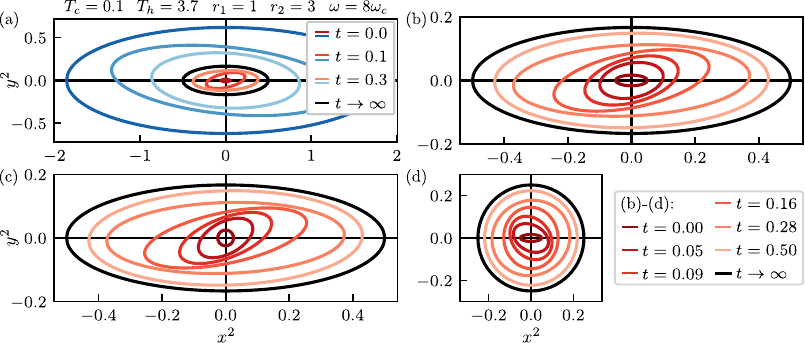}
\caption{\blue{
(a,b) Covariance ellipses from Eq.~\eqref{SM Lyapunov solution} for heating and cooling for the process as in Fig.~\ref{FgSM1} and Fig.~2 in the Letter. (c) As in (b) but with initial condition with $r_1=r_2=2$. (d) As in (b) but with process (but not initial condition) defined with $r_1=r_2=2$.}}\label{FgSM2} 
\end{figure}

In Fig.~\ref{FgSM3} we further illustrate the relation between the
direction of rotation and the shape of covariance ellipses. As
explained above and in Fig.~2f in the Letter, a more circular (less
eccentric) ellipse [sign$(3-$ratio$)$=1] leads to a surprising
counter-clockwise rotation [sign$($angle-change$)$=1]. The overlap of
the curves in Fig.~\ref{FgSM3}c,d corroborates this explanation. Small
deviations between the curves emerge since the heuristic explanation only
applies to ellipses with angle$(t)=0$. 
\begin{figure}[ht!!]
\centering
\includegraphics[width=1.\textwidth]{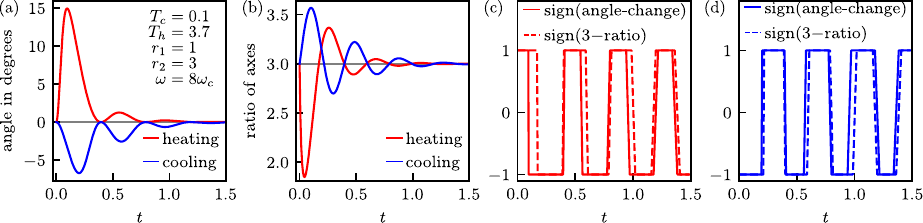}
\caption{\blue{
(a)~Same as in 2e in the Letter. (b) Ratio of the axes of the
    covariance ellipse. Values below $3$ reflect more circular (less
    eccentric) ellipses compared to the initial condition and the
    steady state. (c,d) Direction of rotation (clockwise rotation is
    $+1$) and indicator of shape ($\pm 1$ means more/less round, i.e.,
    less/more eccentric) for heating (c) and cooling (d).}} 
\label{FgSM3}
\end{figure}

In Fig.~\ref{FgSM4} we repeat the presentation of Fig.~\ref{FgSM3} for
a case where the eigenvalues of the drift matrix are real (see Fig.~2d
in the Letter for $\om<\om_c$).  We observe that (opposite) rotational
motions also occur for the case of real eigenvalues, which illustrates
that (effective) rotational motions do \emph{not only} emerge for
complex eigenvalues.  A difference with respect to Fig.~\ref{FgSM3} is
that the angles do not cross $0$ such that the explanation for the
overlaps in Fig.~\ref{FgSM4}c,d only applies at $t=0$. 
\begin{figure}[ht!!]
\centering
\includegraphics[width=1.\textwidth]{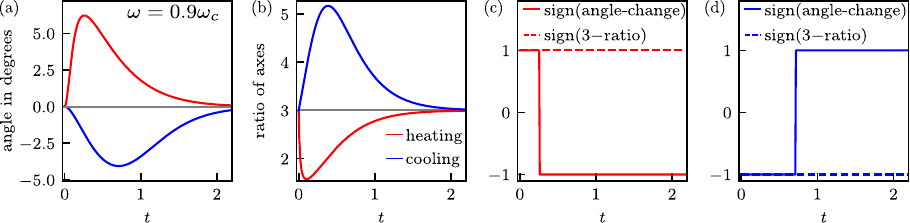}
\caption{\blue{As in Fig.~\ref{FgSM3} but for $\om=0.9\om_c$, i.e.,\ eigenvalues of the drift matrix $\f A$ are real (see Fig.~2d in the Letter).}}\label{FgSM4}
\end{figure}

\subsection*{Relevance and generality of the observation of counterintuitive rotations}
So far we only investigated the origin of the counterintuitive
rotations in the two-dimensional example.  However, since such
counterintuitive rotations already occur in this linear,
low-dimensional example, it is to be expected that such motions also
occur for more general driven systems.  In particular, if
two-dimensional subspaces are described by the example above one
immediately has this rotation in the subspace of the more general
dynamics. Generally, one expects opposite rotations during heating and
cooling (and therefore one of the two has to rotate opposite to the
driving) due to the difference in sign of $\bSig_{{\rm
    s},i}-\bSig_{{\rm s},w}=(T_i/T_w-1)\bSig_{{\rm s},w}$ for $i=h,c$
in Eq.~\eqref{SM Lyapunov solution} [Eq.~(4) in the Letter] as pointed out above. 

The relevance of this observation is twofold. On the one hand, it
further emphasizes the asymmetry between heating and cooling, and that
the process does \emph{not} pass through locally equilibrated states
(i.e., the system \emph{cannot} be described by a time-dependent
temperature). On the other hand, it is also relevant for general
relaxation phenomena, i.e.\ beyond thermal relaxation. For example,
imagine one observes the part of the relaxation process in
Fig.~\ref{FgSM1}a,b for $t\in[0,0.1]$. If one only observes the
apparent counterclockwise rotation of the probability density, one
would never guess that the underlying driving is actually in the
clockwise direction. Therefore, awareness of this counterintuitive
phenomenon might prove useful to avoid false conclusions; and a deep
understanding of this phenomenon helps to arrive at correct
conclusions. 
}

\section{Consistent comparison of equilibrium and non-equilibrium steady states}
We here discuss under which circumstances we consider a comparison of
equilibrium (EQ) and non-equilibrium steady states (NESS), or of different
NESS, to be \emph{consistent}.

In short, we consider a comparison to be
consistent if tuning the driving strength does \emph{not} change the
steady-state density. Before we explain this in detail, we want to
stress that a consistent comparison \emph{is by no means required for the
statement of the thermal relaxation asymmetry to be valid}, since this
statement is proven for \emph{any} NESS with linear drift in the
Letter. Therefore, we were able to chose the physical example of a
Rouse chain in a shear flow to illustrate the relaxation asymmetry in
Fig. 1 in the Letter (which in fact does \emph{not} represent a consistent
comparison). The consistent comparison is, however, necessary for the
statement of ``accelerated relaxation'' since this statement compares
the relaxation speed towards an NESS with the relaxation speed in the
\emph{corresponding} passive system relaxing into an equilibrium
steady state. 

In the Letter, we use Eq.~(3), i.e. Eq.~\eqref{SM Lyapunov ss}, to
obtain the decomposition $\f A=(\f D_i+\balpha_i)\bSig_{{\rm
    s},i}^{-1}$ [Eq.~(7) in the Letter] with $\balpha_i^T=-\balpha_i$
for the linear drift matrix $\f A$. Note that here $\f
D_i,\balpha_i,\bSig_{{\rm s},i}\propto T_i$ all increase linearly with
temperature but the product $\f A$  involving $\bSig_{{\rm
    s},i}^{-1}\propto T_i^{-1}$ is temperature independent. Any $\f A$
from this decomposition fulfills Eq.~(3) in the Letter,
i.e. Eq.~\eqref{SM Lyapunov ss}, with the given $\bSig_{{\rm s},i}$,
and in turn any $\f A$ implying a steady-state covariance $\bSig_{{\rm
    s},i}$ via Eq.~(3) in the Letter can be decomposed with this
$\bSig_{{\rm s},i}$ according to $\f A=(\f D_i+\balpha_i)\bSig_{{\rm
    s},i}^{-1}$. 
%
%
%
The advantage of the latter form is that it allows to systematically
compare NESS dynamics (or in the special case reversible dynamics)
$\rmd\x_t=-\f A\x_t\rmd t+\bsig\rmd\f W_t$ with different $\f A$ that
possess different driving strengths but the same steady-state
density. This comparison is performed by tuning the parameter
$\alpha_i$ (reversible systems are obtained by setting $\alpha_i=\f 0$) for a
given $\bSig_{{\rm s},i}$, which then yields $\f A$ via $\f A=(\f
D_i+\balpha_i)\bSig_{{\rm s},i}^{-1}$ [Eq.~(7) in the Letter]. We
consider such a comparison to be \emph{consistent}, in contrast to a
comparison where tuning the irreversible driving alters $\bSig_{{\rm
    s},i}$ and thus the steady-state density.  

An example for a driving that does \emph{not} yield a consistent
comparison is the shear flow in Fig. 1 in the Letter and in
Eqs.~\eqref{Rouse first equation}-\eqref{Rouse last equation}. We
discuss this comparison in detail now. For simplicity we consider a
single particle $N=1$ in the $x$-$y$ plane subject to the confining
potential and shear flow [see 
Eqs.~\eqref{Rouse C}-\eqref{Rouse    rotation}]
described by the equation of motion $\rmd\x_t=-\f
A\x_t\rmd t+\sqrt{2}\rmd\f W_t$ with drift matrix 
\begin{align}
\f A=\begin{bmatrix}\cos(\theta)&-\sin(\theta)\\\sin(\theta)&\cos(\theta)\end{bmatrix}\begin{bmatrix}r_x&0\\0&r_y\end{bmatrix}\begin{bmatrix}\cos(\theta)&\sin(\theta)\\-\sin(\theta)&\cos(\theta)\end{bmatrix}+\begin{bmatrix}0&\omega\\0&0\end{bmatrix}.\label{rotated sheared 2d drift}
\end{align}
This drift originates from a (rotated) confining potential with
confinement strength quantified by $r_x,r_y>0$, plus a shear flow of
strength $\omega$ (both exactly as shown in Fig.~1a in the
Letter). The drift without the shear flow $\om=0$ is symmetric and
therefore gives rise to reversible dynamics with steady-state
covariance [see Eq.~(3) or (7) in the Letter for $\f D=\mathbbm 1$] 
\begin{align} 
\bSig_{\rm s}=\begin{bmatrix}\cos(\theta)&-\sin(\theta)\\\sin(\theta)&\cos(\theta)\end{bmatrix}\begin{bmatrix}1/r_x&0\\0&1/r_y\end{bmatrix}\begin{bmatrix}\cos(\theta)&\sin(\theta)\\-\sin(\theta)&\cos(\theta)\end{bmatrix}.\label{rotated Sigma} 
\end{align}
The shear flow $\om\ne0$ renders the dynamics irreversible. However,
since now it is not of the form $\balpha\bSig_{\rm s}^{-1}$ with $\balpha^T=-\balpha$ as in Eq.~(7) in the Letter, the steady-state covariance for $\om\ne0$ will no longer be given by Eq.~\eqref{rotated Sigma}, i.e.\ the steady-state Lyapunov equation [see Eq.~(3) in the Letter or Eq.~\eqref{SM Lyapunov ss}]
for $\f A$ with $\omega\ne0$ will give rise to another steady-state
different from Eq.~\eqref{rotated Sigma} which corresponds to $\omega=0$. Therefore, comparing systems with different $\omega$ will generally not be \emph{consistent} [opposed a comparing systems with different $\balpha_i$ in Eq.~(7) in the Letter]. 

We illustrate this \emph{inconsistent} comparison by three different
examples. Choosing the parameters $r_x=1,\ r_y=0.1,\ \omega=3,
\theta=-10^\circ$ as in Fig.~1a in the Letter, the eigenvalues of $\f
A$ are $0.55\pm 0.51\rmi$, i.e. compared to $\omega=0$ with
eigenvalues $r_{x,y}$ the statement of faster relaxation as quantified
in Eq.~(8) in the Letter does still hold true, even though the proof
does not apply here (see also Fig.~1b in the Letter where the curves
with the shear flow decay faster at long times). However, if one
instead takes $\omega=0.5, \theta=10^\circ$ the eigenvalues of $\f A$
are $1.08$ and $0.02$ i.e.\ the limiting relaxation is \emph{slower}
compared to the reversible system since $0.02<r_{x,y}$. Thus the
statement of faster relaxation does not apply since the effect of the
shear flow on the steady state is too large. Even more extreme is the
case $\omega=3, \theta=10^\circ$ where the eigenvalues are $1.365$ and
$-0.265$ where the negative eigenvalue implies that the shear flow
destroyed the confining potential in the sense that the resulting
drift no longer corresponds to a confined process.  \emph{This means that this process no longer relaxes into an NESS.} This can, of course, not happen for a consistent comparison since changing only $\balpha_i$ in Eq.~(7) in the Letter does not change the confinement.

\section{Adiabatic entropy production}
The adiabatic entropy production is the housekeeping heat divided by the reservoir temperature and is given by \cite{SM_VandenBroeck2010PRE}
\begin{align}
\dot S_a(t)=\int\rmd\x P(\x,t)\f a_{\rm irr}(\x)^T\f D_i^{-1}\f a_{\rm irr}(\x),
\end{align}
where the irreversible drift in the linear case considered in the Letter reads $\f a_{\rm irr}(\x)=-\f A_{\rm irr}\x=\f -\balpha_i\bSig_{{\rm s},i}^{-1}\x$. Thus, we see that the adiabatic entropy production term scales linearly with $\balpha_i^T\f D_i^{-1}\balpha_i$ as mentioned in the Letter, i.e.\ it scales quadratically in the driving strength.

\section{Long-time scaling of the Kullback-Leibler divergence}
In terms of the eigenvalue $\lambda_1$ of $\f A$ that has the smallest
real part, we know that asymptotically for large $t$ the magnitude of
$\rme^{-\f At}$ is determined by $\rme^{-\Re(\lambda_1)t}$ [there may
  still be oscillations (see Fig.~2h in the Letter) and if $\f A$ is
  not diagonalizable there may also be terms
  $t^k\rme^{-\Re(\lambda_1)t}$ with $k\in\N$ entering, which
  nonetheless are dominated by $\rme^{-\Re(\lambda_1)t}$ for
  sufficiently large $t$]. Note that $\rme^{-\f
  At}\sim\rme^{-\Re(\lambda_1)t}$ implies, via \blue{Eq.~(6) in the Letter}, that $\f X(t)\sim\rme^{-2\Re(\lambda_1)t}$ for
$t\to\infty$. Recall Eq.~(5) in the Letter, i.e. $2\mathcal
D_t^i=\,\tr[\delta\T_i\f X(t)]-\ln\det[\mathbbm 1+\delta\T_i\,\f
  X(t)]$. Considering $\delta\T_i\,\f X(t)=\rme^{-2\Re(\lambda_1)t}\f
M$ for some matrix $\f M$ for large enough $t$ and using that around
$\rme^{-2\Re(\lambda_1)t}\to 0$, we have that $\det[\mathbbm
  1+\rme^{-2\Re(\lambda_1)t}\f M]=1+\tr[\rme^{-2\Re(\lambda_1)t}\f
  M]+\mathcal O[\rme^{-4\Re(\lambda_1)t}]$, and we obtain $\mathcal D_t^i=\mathcal O[\rme^{-4\Re(\lambda_1)t}]$ as illustrated in Fig.~2h in the Letter. This confirms that the limiting relaxation speed is dictated by $\Re(\lambda_1)$, i.e.\ by the smallest real part of eigenvalues of $\f A$. In the reversible case we have $\Re(\lambda_1)=\mu_1$ with the notation in the Letter, and $\mu_1=r_1$ for the example considered in Fig.~2 in the Letter.

Note that we did not formally exclude the case that the order
$\rme^{-4\Re(\lambda_1)t}$ also vanishes; in this situation we would need to consider even higher orders. It is likely that this case can be generally excluded, however, since no results hinge on the specific scaling, we do not go into more detail here.

\section{Effective stiffness}
The effective stiffness $\hat{r}_j(\omega)\equiv-\ln(x_j^t)/2t$ [such
  that $x_j^t=\rme^{-2\hat r_j(\omega)t}$] is defined as the stiffness
of the confining potential of a reversible system that has the same
thermal relaxation properties as the considered system, where $x_j^t$
for $j=1,\dots,d$ are the eigenvalues of the matrix $\f
X(t)\equiv\rme^{-\f A t}\bSig_{{\rm s},w}\rme^{-\f A^T t}{\bSig_{{\rm
      s},w}^{-1}}=\rme^{-\f A t}\rme^{-\f A_{-\balpha} t}$ [see
  Eqs.~(6) and (10) in the Letter]. 
In Fig.~3b in the Letter we show $\hat{r}_j(\omega)$ for $j=1,2$ for the two-dimensional system as shown in Fig.~2 in the Letter with driving strength $\omega$. The eigenvalues $x_{1,2}^t$ at $T_w=1$ (i.e.\ $T_i$ are measured in units of $T_w$) for this example are computed from 
\begin{align}
\f A&\equiv\begin{bmatrix}r_1&-r_2\om\\r_1\om &r_2\end{bmatrix},\qquad
\bsig=\sqrt 2\mathbbm 1,\qquad
\bSig_{\rm s}=\begin{bmatrix}1/r_1&0\\0&1/r_2\end{bmatrix},\nonumber\\
M&\equiv\sqrt{(r_1-r_2)^2-4r_1r_2\om^2}\in\C,\nonumber\\
\exp(-\f At)&=\frac{\exp[-(r_1+r_2)t/2]}{M}\left[\begin{matrix}M \cosh\left(\frac{Mt}{2}\right) - \abs{r_1-r_2} \sinh\left(\frac{Mt}{2}\right) & 2 \omega r_{2} \sinh\left(\frac{Mt}{2}\right)\\- 2 \omega r_{1} \sinh\left(\frac{Mt}{2}\right) & M \cosh\left(\frac{Mt}{2}\right) + \abs{r_1-r_2} \sinh\left(\frac{Mt}{2}\right)\end{matrix}\right],\nonumber\\
x_1^tx_2^t&=\det[\f X(t)]=\exp[-2(r_1+r_2)t],\nonumber\\
x_1^t+x_2^t&=\tr[\f X(t)]=2\exp[-(r_1+r_2)t]\left[1+2\frac{(r_1-r_2)^2}{M^2}\sinh^2\left(\frac{Mt}{2}\right)\right]\nonumber\\
x_{1,2}^t&=\frac{\tr(\f X(t))}{2}\pm\sqrt{\frac{\tr^2(\f X(t))}{4}-\det(\f X(t))}.
\end{align}

\section{Log-norm inequality}
In the Letter we use the log-norm inequality $\norm{\exp(\f Mt)}\le\exp[\mu(\f M)t]$ \cite{SM_Dahlquist1958} where the log norm is defined via the matrix norm $\norm{\f M}\equiv\sup_{\f v\in\R^d\setminus\f 0}\norm{\f M\f v}_2/\norm{\f v}_2$ (also known as operator norm) where $\norm{\f v}_2=\sqrt{\f v^T\f v}$ as 
\begin{align}
\mu(\f M)\equiv\lim_{h\to0^+}\frac{\norm{\mathbbm 1+h\f M}-1}{h}.\label{log norm def} 
\end{align}
Writing the matrix norm $\norm{\f M}$ in the form $\norm{\f M\f
  v}_2/\norm{\f v}_2=\sqrt{\f v^T\f M^T\f M\f v/\f v^T\f v}$ one sees
that $\norm{\f M}$ is given by the square root of the largest
eigenvalue of the symmetric matrix $\f M^T\f M$. Splitting $\f M=\f
M_s+\f M_a$ with $\f M_s\equiv(\f M+\f M^T)/2=\f M_s^T$ and $\f
M_a\equiv(\f M-\f M^T)/2=-\f M_a^T$ we find that 
\begin{align}
(\mathbbm 1+h\f M)^T(\mathbbm 1+h\f M)&=(\mathbbm 1+h\f M_s-h\f M_a)(\mathbbm 1+h\f M_s+h\f M_a)=1+2h\f M_s+\mathcal{O}(h^2)\nonumber\\
&=(\mathbbm 1+h\f M_s)^T(\mathbbm 1+h\f M_s)+\mathcal{O}(h^2).
\end{align}
This implies that $\norm{\mathbbm 1+h\f M}=\norm{\mathbbm 1+h\f M_s}$
and via Eq.~\eqref{log norm def} that the log norm is solely
determined by the symmetric part $\mu(\f M)=\mu(\f M_s)$. Intuitively
this states that asymmetric contributions (which account for rotations
after exponentiation) do not enter the absolute value in the
exponential bound $\norm{\exp(\f Mt)}\le\exp[\mu(\f M)t]$, which makes
the log norm very useful for our theory. 

From this insight we immediately compute the result used in the Letter, i.e. we use that $-\bbeta\f D_w\bbeta$ with eigenvalues $-\mu_1>-\mu_2>\dots$ is the symmetric part of $-\bbeta\f A\bbeta^{-1}$ to obtain
\begin{align}
\mu(-\widetilde{\f A})\equiv\mu(-\bbeta\f A\bbeta^{-1})=\mu(-\bbeta\f D_w\bbeta)=-\mu_1.
\end{align}

\end{document}